\documentclass[aps,amsmath,amssymb,reprint,nofootinbib]{revtex4-1}

\usepackage{graphicx}
\usepackage{dcolumn}
 \AtBeginDocument{\usepackage{booktabs}}
\usepackage{units}
\usepackage{hyperref}

\newcommand{\ionm}[2]{${}^{#1}$#2${}^+$}
\newcommand{\ion}[1]{#1${}^+$}

\begin{document}

\preprint{}

\title[]{Probing Time Dilation in Coulomb Crystals in a high-precision Ion Trap}

\pacs{37.10.Ty, 06.30.Ft}

\author{J.~Keller}
\affiliation{Physikalisch-Technische Bundesanstalt, Bundesallee 100, 38116 Braunschweig, Germany}
\author{D.~Kalincev}
\affiliation{Physikalisch-Technische Bundesanstalt, Bundesallee 100, 38116 Braunschweig, Germany}
\author{T.~Burgermeister}
\affiliation{Physikalisch-Technische Bundesanstalt, Bundesallee 100, 38116 Braunschweig, Germany}
\author{A.~P.~Kulosa}
\affiliation{Physikalisch-Technische Bundesanstalt, Bundesallee 100, 38116 Braunschweig, Germany}
\author{A.~Didier}
\affiliation{Physikalisch-Technische Bundesanstalt, Bundesallee 100, 38116 Braunschweig, Germany}
\author{T.~Nordmann}
\affiliation{Physikalisch-Technische Bundesanstalt, Bundesallee 100, 38116 Braunschweig, Germany}
\author{J.~Kiethe}
\affiliation{Physikalisch-Technische Bundesanstalt, Bundesallee 100, 38116 Braunschweig, Germany}
\author{T.~E.~Mehlst\"aubler}
\affiliation{Physikalisch-Technische Bundesanstalt, Bundesallee 100, 38116 Braunschweig, Germany}

\date{\today}

\begin{abstract}
Trapped-ion optical clocks are capable of achieving systematic fractional frequency uncertainties of $10^{-18}$ and possibly below. However, the stability of current ion clocks is fundamentally limited by the weak signal of single-ion interrogation. We present an operational, scalable platform for extending clock spectroscopy to arrays of Coulomb crystals consisting of several tens of ions, while allowing systematic shifts as low as $10^{-19}$. Using a newly developed technique, we observe 3D excess micromotion amplitudes inside a Coulomb crystal with atomic spatial resolution and sub-nanometer amplitude uncertainties. We show that in ion Coulomb crystals of $\unit[400]{\mu m}$ and $\unit[2]{mm}$ length, time dilation shifts of \ion{In} ions due to micromotion can be close to $1\times10^{-19}$ and below $10^{-18}$, respectively. In previous ion traps, excess micromotion would have dominated the uncertainty budget for spectroscopy of even a few ions. By minimizing its contribution and providing a means to quantify it, this work opens up the path to precision spectroscopy in many-body ion systems, enabling entanglement-enhanced ion clocks and providing a well-controlled, strongly coupled quantum system.
\end{abstract}

\maketitle
\section{Introduction}
Optical atomic clocks \cite{Ludlow2014} are currently the most accurate man-made devices, capable of providing frequency references with fractional uncertainties approaching one part in $10^{18}$. The two most successful approaches thus far are ensembles of $10^3$ to $10^4$ neutral atoms stored in an optical lattice (e.g.~\cite{McGrew2018,Lisdat2016,Nemitz2016,Campbell2017}), and single trapped ions (e.g.~\cite{Chou2010a,Dube2014,Huntemann2016}). The latter benefit from the ability to be strongly confined due to their charge, which, to first order, does not affect the electronic energy structure (by definition, the time-averaged electric field vanishes at the equilibrium ion position). The resulting strong localization and excellent control over the trapping environment for single ions are an ideal premise for high-accuracy spectroscopy. Besides being excellent candidates for optical clocks, ions also provide transitions with high sensitivities for tests of general relativity \cite{Pruttivarasin2015,Dzuba2016}, or the search for physics beyond the standard model \cite{Rosenband2008,Huntemann2014a,Godun2014}. Stable and reproducible optical clocks also give rise to novel applications like chronometric leveling, which uses relativistic time dilation to determine differences in gravitational potential \cite{Bjerhammar1985, Vermeer1983}. With fractional frequency resolutions of $1\times10^{-18}$, corresponding to $\unit[1]{cm}$ of height difference on Earth's surface, optical clocks become competitive sensors for geodesy \cite{Lion2017,Mueller2017,Mehlstaeubler2017}.

However, current single-ion clocks are fundamentally limited by their statistical uncertainty due to the low signal-to-noise ratio of a single quantum absorber \cite{Itano1993}. The corresponding instabilities of a few $10^{-15}/\sqrt{\tau}$ necessitate measurement times $\tau$ of tens of days before resolving the atomic transition frequency to within $10^{-18}$. These long integration times complicate comparisons between clocks and significantly impede practical applications. So far, it has been unclear whether precision spectroscopy can be extended to many ions for providing fast and ultra-stable clock spectroscopy \cite{Champenois2010,Herschbach2012,Arnold2015}, without compromising accuracy. Here, we demonstrate the possibility to scale the high level of control from a single ion to spatially extended and strongly coupled many-body systems with corresponding systematic uncertainties as low as $1\times10^{-19}$. A clock based on $N$ ions can reach a given instability $N$ times faster, and further allows the implementation of novel clock schemes, such as cascaded interrogation of separate ensembles \cite{Rosenband2013,Borregaard2013} or exploiting quantum correlations \cite{Leroux2010,Kessler2014,Lebedev2014}. The unperturbed storage of extended multi-ion ensembles demonstrated here is also a key requirement for scaling ion based quantum simulation \cite{Schneider2012,Blatt2012,Hess2017} and quantum information processing \cite{Blatt2008} systems to sizes and computation times at which they outperform classical computations.

In this work, we present the operation of a new platform for precision spectroscopy. In previous clock ion traps, micromotion-induced time dilation shifts increased by $10^{-17}$ over $\unit[3]{\mu m}$ \cite{Chou2010a} (or $10^{-18}$ over $\unit[20]{\mu m}$ \cite{Keller2016}). Our platform is capable of storing $\unit[400]{\mu m}$-long ion chains with shifts close to $1\times10^{-19}$. This is shown using a new measurement technique which images time dilation shifts with $10^{-20}$ resolution for each individual ion within a Coulomb crystal. Since these findings resolve one of the major concerns in precision spectroscopy with ion ensembles, they are a crucial prerequisite for future clocks based on this approach. In combination with a detailed analysis of further systematic shifts \cite{Companion}, they allow us to show that the overall contribution of multi-ion operation to the frequency uncertainty of an \ion{In} clock can be below $1\times10^{-19}$. Our approach supports ion numbers on the order of $100$ at this level of accuracy. The scalability of our ion trap platform is also useful for cascaded clock operation and storage of quantum information.

\section{Ion trap array}

\begin{figure}
  \centerline{\includegraphics[width=.48\textwidth]{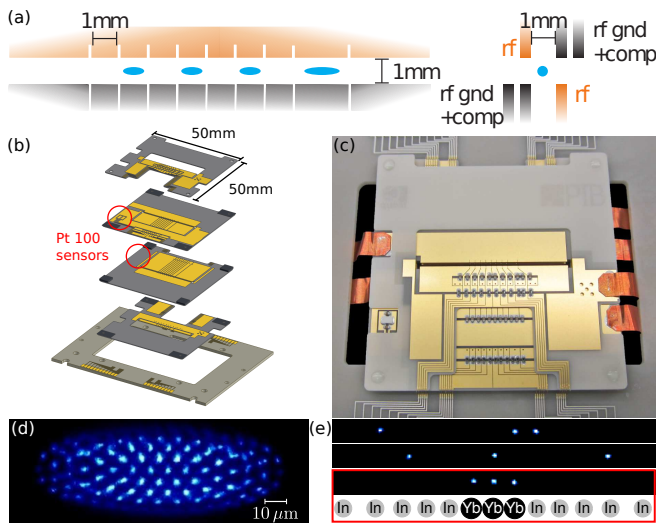}}
  \caption{\label{trapfigure}Precision scalable ion trap array. (a) Electrode geometry: the trap consists of one $\unit[2]{mm}$-long and seven $\unit[1]{mm}$-long segments. (b) Trap assembly from a stack of four wafers. (c) Photograph of the trap, showing the onboard low-pass filters and one thermistor. (d) Large Coulomb crystal (here: \ion{Yb} ions), which can serve as a high-stability frequency reference. (e) Linear \ion{In} (dark) / \ion{Yb} (fluorescing) crystals. The bottom configuration is chosen for high-accuracy clock operation. Its axial length is $\unit[219]{\mu m}$ ($\unit[61]{\mu m}$) for $\nu_\mathrm{ax}=\unit[30]{kHz}$ ($\nu_\mathrm{ax}=\unit[205]{kHz}$), the parameters used in \cite{Companion}.}
\end{figure}

Our platform consists of a linear rf trap array (shown in Fig~\ref{trapfigure}a-c), developed for precision spectroscopy on separate ion ensembles. In this type of trap, large Coulomb crystals can be stored to serve as high-stability frequency references, while short chains on the order of ten ions allow clock operation with high accuracy (see Fig.~\ref{trapfigure}d-e). Separation into relatively small linear crystals limits the complexity of the motional spectrum, provides approximately homogeneous environmental conditions and supports internal state readout with single-ion resolution, while further scaling is achieved by the storage of multiple such ensembles. The trap array consists of four AlN wafers with gold electrodes, the geometry of which is based on the calculations in Ref.~\cite{Herschbach2012}. Assembly from a small number of laser-cut monolithic wafers with laser-structured gold layers ensures scalability and symmetrically shaped electrodes with manufacturing tolerances below $\unit[10]{\mu m}$. Integrated Pt100 sensors allow for in-situ temperature measurements during clock operation. Further trap features include 3D optical access, dedicated electrodes for 3D stray field compensation in each segment, and integrated RC filters close to each electrode to prevent pickup electronic noise from exciting eigenmodes of the Coulomb crystals. The electrical design has previously been tested in a simple prototype trap \cite{Pyka2014}.

\section{Time dilation in ion Coulomb crystals}

The strongest concern for precision spectroscopy in extended ion many-body systems is the time dilation due to fast motion of individual ions, driven by the confining rf field at $\Omega_\textrm{rf}$ (excess micromotion, EMM). Our concept relies on the storage of all ions at positions of minimal rf field amplitude, and is therefore not restricted to atomic species that allow for a cancellation of micromotion induced AC Stark and time dilation shifts \cite{Madej2012}. This choice also avoids ion heating by an rf enabled coupling of noise at frequencies of $\Omega_\mathrm{rf}\pm\omega_i$ to motional modes at $\omega_i$ \cite{Blakestad2009}. However, while a single ion can be placed at the point of lowest (ideally, vanishing) rf field amplitude by the application of static electric fields (see, e.g. \cite{Huber2014,Keller2015,Gloger2015,Meir2017}), doing so with a chain of ions relies on the existence of a true nodal line in the trapping field. Axial rf field components prevent the formation of such a line when the translation symmetry of the trap is broken by its finite length, segmentation, and possible manufacturing imperfections. Secondly, the line of minimal rf amplitude is not necessarily identical to the line of minimal radial potential and it is not clear that the available compensation fields can correct for this over the whole extent of the crystal. Finally, collective effects exist within the crystal, in which ions are affected by the fluctuating Coulomb repulsion from neighboring ions undergoing micromotion \cite{Landa2012,Arnold2015}. We show in Appendix \ref{collectiveEMM} that this last effect scales as $(\omega_\mathrm{ax}/\Omega_\mathrm{rf})^2$ in linear ion chains at an axial trapping frequency $\omega_\mathrm{ax}$ and is negligible for our parameters.

\begin{figure}
\centerline{\includegraphics[width=.5\textwidth]{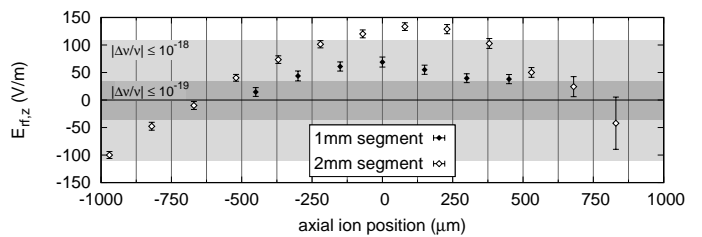}}
\caption{\label{PMTresults}Single-ion measurements of the axial rf field component in two trap segments of $\unit[1]{mm}$ and $\unit[2]{mm}$ length, respectively, at an rf amplitude of $U_\mathrm{rf}\approx\unit[800]{V}$. The shaded areas correspond to time dilation shifts below $10^{-18}$ ($10^{-19}$) for \ionm{115}{In}.}
\end{figure}

\subsection{Single-ion time dilation measurements}

Using a photomultiplier tube (PMT) and the photon-correlation technique \cite{Berkeland1998}, we have previously demonstrated the capability of quantifying the frequency shifts resulting from EMM with uncertainties below $10^{-20}$ for a single ion \cite{Keller2015}. Here, using this technique, we first benchmark the new trap array by measuring the micromotion amplitudes $z_{0,\mathrm{emm}}^\mathrm{(Yb)}$ of a single \ionm{172}{Yb} ion at different positions along the trap axis. The observed rf electric field amplitudes $E_\mathrm{rf,z}=m_\mathrm{Yb}\Omega^2_\mathrm{rf}z_{0,\mathrm{emm}}^\mathrm{(Yb)}/e$ are below $\unitfrac[80]{V}{m}$ ($\unitfrac[140]{V}{m}$) over an entire $\unit[1]{mm}$-long ($\unit[2]{mm}$-long) segment, as shown in Fig.~\ref{PMTresults}. The corresponding fractional time dilation shifts for \ionm{115}{In}, $\vert\Delta\nu/\nu_0\vert=(\Omega_\mathrm{rf}z_{0,\mathrm{emm}}^\mathrm{(In)})^2/(2c^2)=(\Omega_\mathrm{rf}z_{0,\mathrm{emm}}^\mathrm{(Yb)})^2/(2c^2)(m_\mathrm{In}/m_\mathrm{Yb})^2$, are below $5.3\times10^{-19}$ ($1.6\times10^{-18}$). In these relations, $m_\mathrm{Yb}$ and $m_\mathrm{In}$ denote the respective ion mass, $c$ the speed of light, and $e$ the elementary charge (see Appendix \ref{EMMcs} for a comprehensive list of relations between measures of EMM). The change in time dilation shift over a characteristic ion distance of $\unit[10]{\mu m}$ is below $4\times10^{-20}$ everywhere in these segments, with a gradient three orders of magnitude smaller than has previously been reported \cite{Chou2010a}. The resulting axial range with shifts below $10^{-18}$ allows for the storage of more than $50$ ions in each segment.

The question remains whether single-ion measurements can predict the rf driven motion of multiple simultaneously trapped ions. We have therefore developed a method to simultaneously observe EMM in individual ions within a Coulomb crystal.

\subsection{Stroboscopic imaging technique}
\begin{figure}[b]
  \centerline{\includegraphics[width=.5\textwidth]{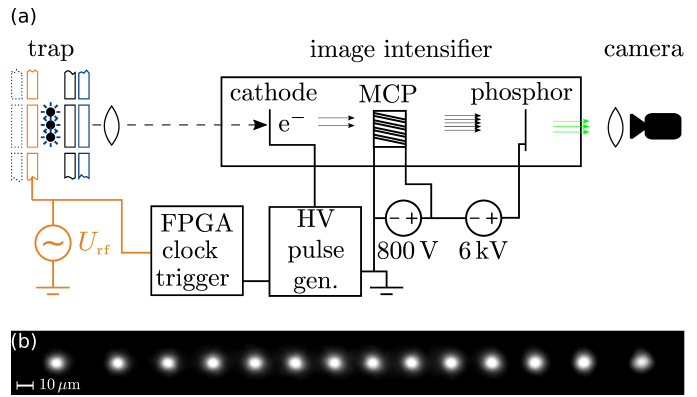}}
  \caption{\label{photon_correlation_II_exp_scheme}(a) Setup for stroboscopic micromotion measurements across Coulomb crystals with atomic resolution. (b) Stroboscopic image of a 14-ion \ionm{172}{Yb} crystal for one rf phase and laser direction setting with a total integration time of $\unit[2700]{s}$.}
\end{figure}

Our experimental setup for spatially resolved micromotion measurements in Coulomb crystals is shown in Fig.~\ref{photon_correlation_II_exp_scheme}a: Ion fluorescence is imaged onto an image intensifier,\footnote{Proxivision BV 2581 TZ 5N} the cathode of which is driven with $\unit[10]{ns}$ square pulses, corresponding to about $1/4$ of the rf period. These pulses are triggered by the trap drive, with a variable delay $\delta t$ to adjust the relative phase. Details on the electronic circuit can be found in Ref.~\cite{Beev2017}. Photo-electrons are subsequently multiplied using a multi-channel plate assembly and converted back to photons by acceleration onto a fluorescent screen. Finally, the screen image is observed on an EMCCD camera and integrated separately for each of four phase settings and three laser directions. All 12 settings are interleaved and the collected images are added up in post-processing. Figure \ref{photon_correlation_II_exp_scheme}b shows an example of such an image with a total exposure of $\unit[2700]{s}$. The spatial resolution is slightly deteriorated by the image intensifier, as the applied cathode pulses of ca.~$\unit[100]{V}$ are half of the design value, allowing for more transverse electron motion during the time of flight. Since the mean photon number during each gate pulse is about $4\times10^{-5}$ per ion, mutual electron repulsion at this stage is negligible. Individual ions can be clearly resolved in the final image, and circular regions of interest are used to separate their signals. No modulation was observed in stray light imaged with this setup, ensuring that there is no additional modulation due to, e.g.~electronic noise at the trap drive frequency.\\

We use \ionm{172}{Yb} ions as probes for the rf electric field. Fluorescence on the ${}^2\mathrm{S}_{1/2}\leftrightarrow {}^2P_{1/2}$ transition at $\unit[370]{nm}$ is modulated by EMM as
\begin{equation}
S(t)=S_0+\Delta S\cos(\Omega_\mathrm{rf} t)\;\textnormal{.}
\label{emmsignal}
\end{equation}
The transfer function of the image intensifier setup can be written as
\begin{equation}
d(t)=\left(\mathrm{rect}\left(\frac{1}{\alpha}\frac{\Omega_\mathrm{rf}}{2\pi}t-\delta t\right)+\xi\left(t-\delta t\right)\right)*\mathrm{comb}\left(\frac{\Omega_\mathrm{rf}}{2\pi}t\right)\;\textnormal{,}
\label{detectionfunction}
\end{equation}
where $\mathrm{rect}$ and $\mathrm{comb}$ are the rectangular and Dirac comb functions, $\alpha<1$ is the fraction of the rf period for which the intensifier is gated, $\xi(t)$ accounts for a possible deviation from ideal rectangular pulses, and $*$ denotes convolution.
The observed signal is the product of Eqns.~\ref{emmsignal} and \ref{detectionfunction}, integrated over the detection time:
\begin{align}
O&=\int_{t_\mathrm{det}}S(t)d(t)dt\\\nonumber
&=S_0\int_{t_\mathrm{det}}d(t)dt+\Delta S\int_{t_\mathrm{det}}\cos(\Omega_\mathrm{rf} t)d(t)dt\\\nonumber
&=S_0\int_{t_\mathrm{det}}d(t)dt+\Delta S\int_{t_\mathrm{det}}\mathrm{Re}\left[e^{i\Omega_\mathrm{rf} t}\right]d(t)dt\\\nonumber
&\approx S_0\cdot D(0)+\Delta S\cdot\mathrm{Re}\left[D(\Omega_\mathrm{rf})\right]\;\textnormal{,}
\end{align}
where the last step assumes $t_\mathrm{det}\gg\frac{2\pi}{\Omega_\mathrm{rf}}$, such that the integration limits can be replaced by $\pm\infty$, $D(\omega)$ is the Fourier transform of $d(t)$,
\begin{align}
D(\omega)&\propto\left(\mathrm{sinc}\left(\pi\alpha\frac{\omega}{\Omega_\mathrm{rf}}\right)+\Xi(\omega)\right)\\\nonumber	
&\times\mathrm{comb}\left(\frac{\omega}{\Omega_\mathrm{rf}}\right)\times e^{i\delta t/\Omega_\mathrm{rf}}\;\textnormal{,}\nonumber
\end{align}
and $\Xi(\omega)$ the Fourier transform of $\xi(t)$. By adjusting $\delta t$, the magnitude of $\mathrm{Re}[D(\Omega_\mathrm{rf})]$ is varied in order to distinguish the $\Delta S$ and $S_0$ contributions:
\begin{align}
O(\delta t)&=S_0\cdot D(0)+\Delta S\left\vert D(\Omega_\mathrm{rf})\right\vert\cos(\Omega_\mathrm{rf}\delta t)\\\nonumber
&\propto(1+\Xi(0))\cdot S_0\\\nonumber
&+(\mathrm{sinc}(\pi\alpha)+\Xi(\Omega_\mathrm{rf}))\cdot\Delta S\cos(\Omega_\mathrm{rf}\delta t)\;\textnormal{.}\nonumber
\end{align}
The quantity $\Delta S/S_0$, which is used to determine the micromotion amplitude \cite{Keller2015}, is thus observed with a contrast of
\begin{equation}
C=\frac{\mathrm{sinc}(\pi\alpha)+\Xi(\Omega_\mathrm{rf})}{1+\Xi(0)}\;\textnormal{.}
\end{equation}
Due to the unknown contribution of $\xi(t)$, $C$ needs to be determined experimentally. We perform this calibration using the long-term stable axial micromotion amplitude of a single ion, as determined in PMT measurements before and after, as a reference (see Ref.~\cite{Beev2017}). The evaluation according to \cite{Keller2015} yields a micromotion modulation index $\beta_i=\vec{k}_i\vec{r}_\mathrm{emm}$, where $\vec{r}_\mathrm{emm}$ is the micromotion amplitude and $\vec{k}$ the laser wave vector. The rf electric field amplitude component along the $i$th laser direction can be inferred from $\beta_i$ via
\begin{equation}
E_{\mathrm{rf},i}=\frac{m\Omega_\mathrm{rf}^2}{\vert \vec{k}_i\vert e}\beta_i=\frac{m\Omega_\mathrm{rf}^2}{e}r_{i,\mathrm{emm}}\;\textnormal{,}
\end{equation}
where $\vec{k}_i$ is the laser wave vector. The resulting time dilation shift depends on $E_\mathrm{rf}$ as
\begin{equation}
\frac{\Delta\nu}{\nu_0}=-\left(\frac{eE_\mathrm{rf}}{2mc\Omega_\mathrm{rf}}\right)^2\;\textnormal{.}
\end{equation}
Conversion relations between these and other commonly used parameters to quantify EMM (such as stray field amplitudes, equilibrium position displacements, etc.) are summarized in Appendix \ref{EMMcs}.

\subsection{Experimental results}

\begin{figure*}
  \centerline{\includegraphics[width=\textwidth]{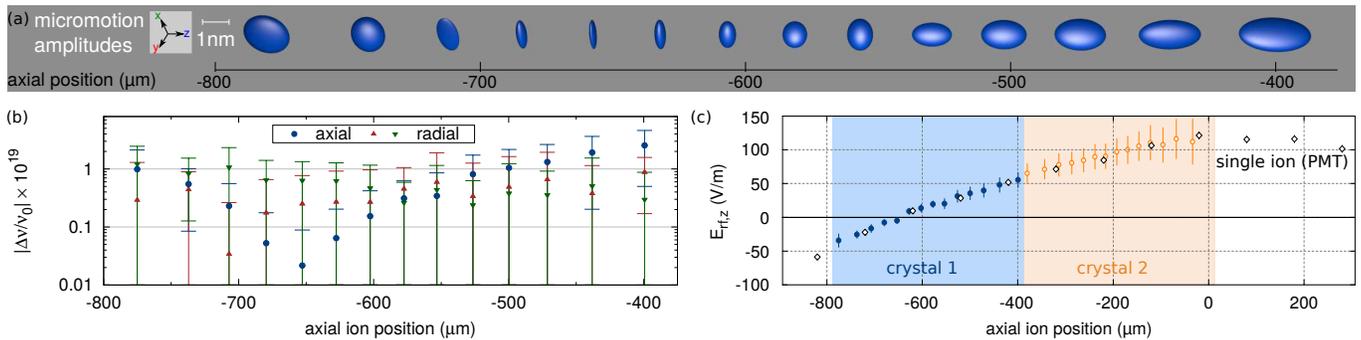}}
  \caption{\label{EMMfigure}Measured excess micromotion in Coulomb crystals. (a) Measured 3D micromotion amplitudes of individual ions inside a Coulomb crystal in the $\unit[2]{mm}$ segment. (b) Time dilation shifts for \ionm{115}{In} ions due to $E_\mathrm{rf}$ at the respective positions. (c) Comparison of the axial rf fields determined in single-ion PMT measurements and by stroboscopic imaging whole Coulomb crystals. Integration times were $\unit[2700]{s}$ ($\unit[900]{s}$) per phase and direction for crystal 1 (2). Sub-figures (a), (b), and ``crystal 1'' in (c) use data from the same measurement (see Appendix \ref{crystal1data} for values and uncertainties).}
\end{figure*}

Figure \ref{EMMfigure}a shows the 3D micromotion amplitudes $\vec{r}_\mathrm{emm}$ measured in a crystal consisting of 14 \ion{Yb} ions, confined at $\nu_\mathrm{ax}=\unit[11.6]{kHz}$. In this figure, the motional amplitudes of the ions have been exaggerated by a factor of $10^4$ with respect to the ion distances, as the largest value is $\vert \vec{r}_\mathrm{emm,Yb}\vert=\unit[1.6]{nm}$. The signals are averaged over $\unit[2700]{s}$ for each of the four phases and three laser directions, and all resulting amplitude uncertainties are below $\unit[0.3]{nm}$. The time dilation shifts due to the mean velocities of $\bar{v}_\mathrm{emm,In}=\Omega_\mathrm{rf}r_\mathrm{emm,Yb}/2\cdot m_\mathrm{Yb}/m_\mathrm{In}$ for \ionm{115}{In} ions are shown in Fig.~\ref{EMMfigure}b. Those due to axial micromotion are close to $10^{-19}$ for crystals up to $\unit[400]{\mu m}$ in length, which, depending on the axial confinement, corresponds to ca.~$10$ to $40$ ions (at $\nu_\mathrm{ax}=\unit[10\;\mathrm{to}\;60]{kHz}$). The radial components are caused by a drift of stray fields during the measurement time of more than 9 hours for that crystal. As they are uniform across the crystal, more careful suppression below $10^{-19}$ is possible during clock operation, e.g. by monitoring the fluorescence of a single ion with a dedicated PMT during the Doppler cooling phases of the clock cycle \cite{Keller2015}. A comparison of the new method with single-ion PMT measurements, shown in Fig.~\ref{EMMfigure}c for two different crystal positions, shows matching results. This proves that in the chosen symmetric electrode geometry, the observation of a single ion is sufficient to predict the EMM amplitudes across a full linear Coulomb crystal at this level of accuracy. Further, the axial components of these measurements agree within their uncertainty over the course of more than 6 months. This means the time consuming micromotion measurement of the full crystal has to be carried out only once to characterize the trap geometry, while fast micromotion detection of a single probe ion is sufficient during the precision spectroscopy.

\section{Conclusion}

In summary, we have demonstrated an operational scalable platform for simultaneous precision spectroscopy on multiple ions with fractional frequency uncertainties at the $10^{-19}$ level, and shown for the first time that rf-driven micromotion can be controlled in spatially extended many-body systems. Micromotion is major uncertainty contribution in optical ion clocks and other precision measurements \cite{Shaniv2018}, leads to increased heating rates \cite{Blakestad2009}, as well as heating and losses in ion/neutral atom hybrid systems \cite{Cetina2012}. A full uncertainty budget for multi-ion clock operation with \ionm{115}{In}, including all trap-related systematic shifts, can be found in \cite{Companion}. Our findings open up the possibility to overcome the fundamental statistical uncertainty limitation of single-ion optical clocks and implement novel clock schemes, e.g.~using quantum correlations in many-ion systems, while allowing systematic uncertainties at this level.

\begin{acknowledgments}
The authors thank M.~Drewsen for helpful discussions on intensifier gating electronics, N.~Beev for developing the circuit, the PTB departments 5.3 and 5.5 for the collaboration on trap fabrication, and S.~A.~King for helpful comments on the manuscript. This work was supported by DFG through grant ME3648/1-1 and SFB 1227 (DQ-mat), project B03.
\end{acknowledgments}

\bibliography{cc_clock_papers}

\begin{appendix}
\section{\label{collectiveEMM}Collective micromotion effects in an ion chain}
\begin{figure}[h]
\centerline{\includegraphics[width=.25\textwidth]{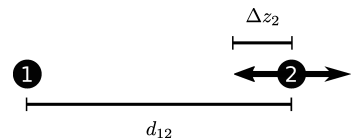}}
\caption{\label{collective_emm_figure}Geometry considered for the derivation of collective micromotion effects in a chain.}
\end{figure}

Consider the situation depicted in Fig.~\ref{collective_emm_figure}: two ions at equilibrium positions $z_1=0$ and $z_2=d_{12}$, of which ion 2 undergoes micromotion due to an rf electric field $E_{\mathrm{rf},2}$ pointing along $z$. The corresponding micromotion amplitude is $\Delta z_2 = eE_{\mathrm{rf},2}/(m\Omega_\mathrm{rf}^2)$. This displacement results in an electric field at the position of ion 1 of
\begin{align}
E_{2\rightarrow1}&=E_1(d_{12}+\Delta z_2)-E_1(d_{12})\\\nonumber
&\approx \left.\frac{\partial E_1}{\partial z_2}\right\vert_{z_2=d_{12}}\Delta z_2\\\nonumber
&=\frac{-e}{4\pi\varepsilon_0}\left.\frac{\partial}{\partial z_2}\frac{1}{z_2^2}\right\vert_{z_2=d_{12}}\frac{eE_{\mathrm{rf},2}}{m\Omega_\mathrm{rf}^2}\\\nonumber
&=\frac{e^2}{2\pi\varepsilon_0}\frac{E_{\mathrm{rf},2}}{m\Omega_\mathrm{rf}^2d_{12}^3}\;\textnormal{,}\nonumber
\end{align}
since $E_1(d_{12})\stackrel{!}{=}0$ at equilibrium positions. Following \cite{James1998}, the equilibrium distances between ions $i$ and $j$ in a harmonically confined chain can be expressed as $d_{ij}=lu_{ij}$, where $u_{ij}$ are of order 1 and independent of trapping parameters, and
\begin{equation}
l^3=\frac{e^2}{4\pi\varepsilon_0m\omega_\mathrm{ax}^2}\;\textnormal{.}
\end{equation}

The relative field experienced by ion 1 thus scales as
\begin{equation}
\frac{E_{2\rightarrow1}}{E_{\mathrm{rf},2}}=\frac{e^2}{2\pi\varepsilon_0}\frac{1}{m\Omega_\mathrm{rf}^2(lu_{12})^3}=\frac{2}{u_{12}^3}\left(\frac{\omega_\mathrm{ax}}{\Omega_\mathrm{rf}}\right)^2\;\textnormal{.}
\end{equation}
Of all the configurations used in our work, the highest value for this expression by far is $1.3\times10^{-3}$, which occurs for the center ions in the crystal of Fig.~\ref{trapfigure}e ($u_{67}=u_{78}\approx0.48$) in the sympathetic cooling scenario of \cite{Companion}, where $\omega_\mathrm{ax}=2\pi\times\unit[205]{kHz}$ and $\Omega_\mathrm{rf}=2\pi\times\unit[24]{MHz}$.

\section{\label{EMMcs}Conversion between common parameters to quantify excess micromotion}
Table \ref{emmcheatsheet} contains the relations between the various quantities used in the literature to quantify excess micromotion. In these relations, $k$ denotes the projection of the laser wave vector onto the micromotion direction, $c$ the speed of light, and $e$ the elementary charge.
\begingroup
\squeezetable
\begin{table*}
\begin{ruledtabular}
\begin{tabular}{lcccccccc}
&$\beta=$&$E_\mathrm{rf}=$&$E_\mathrm{dc}=$&$\Delta x=$&$x_0=$&$v_0=$&$\frac{\Delta\nu}{\nu_0}$&$E_\mathrm{kin}=$\tabularnewline\toprule
$\beta$&&$\frac{m\Omega_\mathrm{rf}^2}{ke}\beta$&$\frac{2m\omega^2}{kqe}\beta$&$\frac{2}{kq}\beta$&$\frac{1}{k}\beta$&$\frac{\Omega_\mathrm{rf}}{k}\beta$&$-\left(\frac{\Omega_\mathrm{rf}}{2ck}\beta\right)^2$&$m\left(\frac{\Omega_\mathrm{rf}}{2k}\beta\right)^2$\tabularnewline\midrule
$E_\mathrm{rf}$&$\frac{ke}{m\Omega_\mathrm{rf}^2}E_\mathrm{rf}$&&$\frac{2\omega^2}{q\Omega_\mathrm{rf}^2}E_\mathrm{rf}$&$\frac{2e}{qm\Omega_\mathrm{rf}^2}E_\mathrm{rf}$&$\frac{e}{m\Omega_\mathrm{rf}^2}E_\mathrm{rf}$&$\frac{e}{m\Omega_\mathrm{rf}}E_\mathrm{rf}$&$-\left(\frac{e}{2cm\Omega_\mathrm{rf}}E_\mathrm{rf}\right)^2$&$m\left(\frac{e}{2m\Omega_\mathrm{rf}}E_\mathrm{rf}\right)^2$\tabularnewline\midrule
$E_\mathrm{dc}$&$\frac{kqe}{2m\omega^2}E_\mathrm{dc}$&$\frac{q\Omega_\mathrm{rf}^2}{2\omega^2}E_\mathrm{dc}$&&$\frac{e}{m\omega^2}E_\mathrm{dc}$&$\frac{qe}{2m\omega^2}E_\mathrm{dc}$&$\frac{qe\Omega_\mathrm{rf}}{2m\omega^2}E_\mathrm{dc}$&$-\left(\frac{qe\Omega_\mathrm{rf}}{4mc\omega^2}E_\mathrm{dc}\right)^2$&$m\left(\frac{qe\Omega_\mathrm{rf}}{4m\omega^2}E_\mathrm{dc}\right)^2$\tabularnewline\midrule
$\Delta x$&$\frac{kq}{2}\Delta x$&$\frac{qm\Omega_\mathrm{rf}^2}{2e}\Delta x$&$\frac{m\omega^2}{e}\Delta x$&&$\frac{q}{2}\Delta x$&$\frac{q\Omega_\mathrm{rf}}{2}\Delta x$&$-\left(\frac{q\Omega_\mathrm{rf}}{4c}\Delta x\right)^2$&$m\left(\frac{q\Omega_\mathrm{rf}}{4}\Delta x\right)^2$\tabularnewline\midrule
$x_0$&$kx_0$&$\frac{m\Omega_\mathrm{rf}^2}{e}x_0$&$\frac{2m\omega^2}{qe}x_0$&$\frac{2}{q}x_0$&&$\Omega_\mathrm{rf}x_0$&$-\left(\frac{\Omega_\mathrm{rf}}{2c}x_0\right)^2$&$m\left(\frac{\Omega_\mathrm{rf}}{2}x_0\right)^2$\tabularnewline\midrule
$v_0$&$\frac{k}{\Omega_\mathrm{rf}}v_0$&$\frac{m\Omega_\mathrm{rf}}{e}v_0$&$\frac{2m\omega^2}{qe\Omega_\mathrm{rf}}v_0$&$\frac{2}{q\Omega_\mathrm{rf}}v_0$&$\frac{1}{\Omega_\mathrm{rf}}v_0$&&$-\left(\frac{1}{2c}v_0\right)^2$&$m\left(\frac{1}{2}v_0\right)$\tabularnewline\midrule
$\frac{\Delta\nu}{\nu_0}$&$\frac{2ck}{\Omega_\mathrm{rf}}\sqrt{\left\vert\frac{\Delta\nu}{\nu_0}\right\vert}$&$\frac{2cm\Omega_\mathrm{rf}}{e}\sqrt{\left\vert\frac{\Delta\nu}{\nu_0}\right\vert}$&$\frac{4mc\omega^2}{qe\Omega_\mathrm{rf}}\sqrt{\left\vert\frac{\Delta\nu}{\nu_0}\right\vert}$&$\frac{4c}{q\Omega_\mathrm{rf}}\sqrt{\left\vert\frac{\Delta\nu}{\nu_0}\right\vert}$&$\frac{2c}{\Omega_\mathrm{rf}}\sqrt{\left\vert\frac{\Delta\nu}{\nu_0}\right\vert}$&$2c\sqrt{\left\vert\frac{\Delta\nu}{\nu_0}\right\vert}$&&$-mc^2\frac{\Delta\nu}{\nu_0}$\tabularnewline\midrule
$E_\mathrm{kin}$&$\frac{2k}{\sqrt{m}\Omega_\mathrm{rf}}\sqrt{E_\mathrm{kin}}$&$\frac{2\sqrt{m}\Omega_\mathrm{rf}}{e}\sqrt{E_\mathrm{kin}}$&$\frac{4\sqrt{m}\omega^2}{qe\Omega_\mathrm{rf}}\sqrt{E_\mathrm{kin}}$&$\frac{4}{q\sqrt{m}\Omega_\mathrm{rf}}\sqrt{E_\mathrm{kin}}$&$\frac{2}{\sqrt{m}\Omega_\mathrm{rf}}\sqrt{E_\mathrm{kin}}$&$\frac{2}{\sqrt{m}}\sqrt{E_\mathrm{kin}}$&$-\frac{1}{mc^2}E_\mathrm{kin}$&\\
\end{tabular}
\end{ruledtabular}
\caption{\label{emmcheatsheet}Conversion between common parameters to quantify excess micromotion: laser modulation index $\beta$, rf electric field amplitude $E_\mathrm{rf}$, electric stray field amplitude $E_\mathrm{dc}$ (component along the direction with ponderomotive confinement at secular frequency $\omega$), ion displacement from the rf node $\Delta x$, micromotion amplitude $x_0$, micromotion peak velocity $v_0$, average time dilation shift $\Delta\nu/\nu_0$ and mean kinetic energy $E_\mathrm{kin}$. A purely ponderomotive confinement with Mathieu parameter $q$ is assumed, such that $\vert q\vert\approx 2\sqrt{2}\omega/\Omega_\mathrm{rf}$.}
\end{table*}
\endgroup

\section{\label{crystal1data}Fig.~\ref{EMMfigure} ``crystal 1'' data}
Table \ref{IIEMMresults_table} contains the results of a stroboscopic micromotion measurement with a 14-ion \ionm{172}{Yb} crystal in the $\unit[2]{mm}$-long segment. This data is shown in Fig.~\ref{EMMfigure} (denoted as ``crystal 1''). Micromotion amplitudes ($x_\mathrm{emm}$, $y_\mathrm{emm}$, $z_\mathrm{emm}$ and $r_\mathrm{emm}=\sqrt{x_\mathrm{emm}^2+y_\mathrm{emm}^2+z_\mathrm{emm}^2}$) are those of the \ionm{172}{Yb} ions, whereas the fractional time dilation shifts $\vert\Delta\nu/\nu_0$ apply to \ionm{115}{In} ions in the presence of the respective rf electric field amplitudes $\vert E_\mathrm{rf}\vert$.\\

\begingroup
\squeezetable
\begin{table*}
\begin{ruledtabular}
\begin{tabular}{llllllllll}
$z$ ($\unit{\mu m}$)&$E_\mathrm{rf,x}$ ($\unitfrac{V}{m}$)&$x_\mathrm{emm}$ ($\unit{nm}$)&$E_\mathrm{rf,y}$ ($\unitfrac{V}{m}$)&$y_\mathrm{emm}$ ($\unit{nm}$)&$E_\mathrm{rf,z}$ ($\unitfrac{V}{m}$)&$z_\mathrm{emm}$ ($\unit{nm}$)&$\vert\vec{E}_\mathrm{rf}\vert$ ($\unitfrac{V}{m}$)&$r_\mathrm{emm}$ ($\unit{nm}$)&$\vert\Delta\nu/\nu_0\vert\times10^{19}$\tabularnewline\toprule
$-775$&$19(16)$&$0.4(4)$&$38(10)$&$0.9(2)$&$-34(10)$&$-0.8(2)$&$55(11)$&$1.3(3)$&$2.5(10)$\\
$-737$&$23(6)$&$0.5(1)$&$32(7)$&$0.8(2)$&$-26(5)$&$-0.6(1)$&$47(6)$&$1.1(1)$&$1.8(5)$\\
$-707$&$6(11)$&$0.2(3)$&$36(11)$&$0.9(3)$&$-17(6)$&$-0.4(1)$&$40(10)$&$1.0(2)$&$1.3(7)$\\
$-680$&$14(10)$&$0.3(2)$&$28(8)$&$0.7(2)$&$-8(5)$&$-0.2(1)$&$32(8)$&$0.8(2)$&$0.9(5)$\\
$-653$&$17(9)$&$0.4(2)$&$28(8)$&$0.7(2)$&$-5(4)$&$-0.1(1)$&$33(8)$&$0.8(2)$&$0.9(4)$\\
$-628$&$18(11)$&$0.4(3)$&$27(7)$&$0.7(2)$&$9(5)$&$0.2(1)$&$34(8)$&$0.8(2)$&$1.0(5)$\\
$-603$&$18(12)$&$0.4(3)$&$24(9)$&$0.6(2)$&$14(6)$&$0.3(1)$&$33(9)$&$0.8(2)$&$0.9(5)$\\
$-578$&$23(8)$&$0.6(2)$&$18(5)$&$0.4(1)$&$19(5)$&$0.5(1)$&$35(6)$&$0.8(2)$&$1.0(4)$\\
$-553$&$27(15)$&$0.6(4)$&$23(9)$&$0.5(2)$&$20(8)$&$0.5(2)$&$41(12)$&$1.0(3)$&$1.4(8)$\\
$-527$&$20(14)$&$0.5(3)$&$17(7)$&$0.4(2)$&$31(9)$&$0.7(2)$&$41(10)$&$1.0(2)$&$1.4(7)$\\
$-500$&$24(14)$&$0.6(3)$&$21(12)$&$0.5(3)$&$35(9)$&$0.8(2)$&$48(11)$&$1.1(3)$&$1.9(9)$\\
$-471$&$28(14)$&$0.7(3)$&$21(8)$&$0.5(2)$&$40(10)$&$1.0(2)$&$53(11)$&$1.3(3)$&$2.3(10)$\\
$-438$&$21(11)$&$0.5(3)$&$25(13)$&$0.6(3)$&$48(11)$&$1.1(3)$&$58(11)$&$1.4(3)$&$2.8(11)$\\
$-399$&$32(7)$&$0.8(2)$&$19(9)$&$0.5(2)$&$55(11)$&$1.3(3)$&$67(10)$&$1.6(2)$&$3.7(11)$\\
\end{tabular}
\end{ruledtabular}
\caption{\label{IIEMMresults_table} Experimental values shown in Fig.~\ref{EMMfigure} a,b, and c (denoted as ``crystal 1'')}
\end{table*}
\endgroup
\end{appendix}

\end{document}